# Cat-Eye Inspired Active-Passive-Composite Aperture-Shared Sub-Terahertz Meta-Imager for Non-Interactive Concealed Object Detection

*Mingshuang Hu[1], Yuzhong Wang\*[1], Zhe Jiang[1], Cheng Pang[1], Ying Li[1], Zhenyu Shao[1], Ziang Yue[1], Yiding Liu[1], Zeming Kong[1], Pengcheng Wang[1], Yifei Wang[1], Axiang Yu[1], Yinghan Wang[1], Wenzhi Li[2], Yongkang Dong[3], Yayun Cheng\*[1], Jiaran Qi\*[1].*

[1]Department of Microwave Engineering, School of Electronics and Information Engineering, Harbin Institute of Technology, Harbin 150001, China.

[2]Department of Microwave Engineering, School of Electronic Information, Hangzhou Dianzi University, Hangzhou 310018, China.

[3]Department of Integrated Circuit Science and Engineering, School of Astronautics, Harbin Institute of Technology, Harbin 150001, China.

E-mail: qi.jiaran@hit.edu.cn(J. Q.), chengyy@hit.edu.cn(Y. C.), hitwyz@stu.hit.edu.cn(Y. W.).

*Abstract*: Within the feline eye, a distinctive tapetum lucidum as a mirror resides posterior to the retina, reflecting the incident rays to simulate light source emission. This secondary emission property enables felines to be highly sensitive to light, possessing remarkable visual capabilities even in dark settings. Drawing inspiration from this natural phenomenon, we propose an active-passive-composite sub-terahertz meta-imager integrated with a bifocus metasurface, a high-sensitivity radiometer, and a low-power signal hidden radiation source. Benefiting from its aperture-shared advantage, this advanced fusion imaging system, enabled to be deployed by a simplified portable hardware platform, allows for the concurrent acquisition of active and passive electromagnetic properties to extend the target detection category and realize multi-mode fusion perception. Notably, it also enables the extraction of radiation and reflection characteristics without additional calibration modules. Experiments demonstrate the multi-target fusion imaging and localized information decoupling with the tailored field of view and emission energy. This compact and multi-mode fusion imaging system may have plenty of potential for airplane navigation positioning, abnormal monitoring, and non-interactive concealed security checks.

**Introduction**

Multi-dimensional electromagnetic (EM) imaging technology reveals hidden target characteristics by skillfully processing EM wave fundamental properties, such as phase, polarization, and spectrum[1-3]. It is extensively applied in non-destructive testing[4,5], stress detection[6], and EM tomography[7]. According to the transmitting-receiving (TR) information mode, multi-dimensional EM imaging systems can be segmented into the active imager (AI), the passive imager (PI), and the active-passive-composite imager (APCI). AI emits modulated signals to the target and then captures the reflected echo, with the problems of specular reflection and speckle interference effect resulting in imaging clarity degradation[8-10]. PI can gather spontaneous emission signals, however, its contrast is constrained in scenarios with minor differences between target and ambient thermal radiation[11-13]. APCI integrates the active emission source and passive detector to obtain simultaneously the radiation and scattering characteristics of targets. This imaging strategy, which combines active-passive independent and complementary EM detection modes while balancing integration, robustness, and adaptability, improves the recognition and anti-interference capability in complicated dynamic scenarios[14].

APCI is mainly implemented through three approaches, hetero-frequency loading, co-frequency time-division loading, and co-frequency concurrent loading. The first mechanism allocates two detection modes across non-overlapping frequency bands, allowing active-passive parallel imaging. This results in higher requirements for antenna bandwidth and cross-talk suppression compared to co-frequency architecture. Furthermore, this frequency division multiplexing (FDM) technology is also accompanied by bulky-size devices[15-17]. The second method maps signals on the identical frequency through time division multiplexing (TDM). This approach offers advantages such as simplified architecture and high-efficiency spectrum utilization, however, it falls short of the capacity to execute real-time output and field of view (FOV) matching[18-20]. Deploying signals on identical frequency bands and implementing spatial interleaving layout, as the third strategy, to accomplish active-passive spatial multiplexing detection[21,22]. This architecture, leveraging concurrent processing across multiple sensing channels, minimizes transmission delays due to increased data throughput while achieving various fusion data deconstruction and multi-mode collaborative detection-making. However, spatial misalignment layout often requires cascading non-spherical lenses, whose sizes exceed other front-end devices, severely limiting integration and expansiveness. Therefore, accomplishing synchronous co-frequency APCI with FOV matching and compact configuration for instantaneous high-resolution APCI remains a critical challenge awaiting a breakthrough.

Metasurfaces, with their robust and flexible EM wave manipulation capabilities, have been deployed into diverse EM detection technologies to tackle sophisticated data, surpassing the confines of conventional optical devices and fostering miniaturization breakthroughs[23-25]. By modulating the EM response of the meta-atoms, they precisely regulate the polarization, frequency, and direction of EM waves at sub-wavelength scales, extensively utilized in domains including holography, stealth, and communication[26-32]. Owing to their lightweight, integrated, and planar characteristics, metasurfaces, as desirable substitutes to traditional optical components, have been implemented in a plethora of compact photonics systems, including hyperspectral cameras, polarization detection, and quantitative phase imaging[33-38]. A single metasurface can integrate multifunctional optical regulation, in particular, adapting to multitasking requirements in miniaturization and lightweight. Up to now, metasurface-based APCI by the directional aperture-shared TR architecture, holding significant potential for breaking through the imaging bottlenecks of FOV mismatching, delayed output, and bulky size, has not been explored[39-42].

In this work, metasurface-based sub-terahertz APCI is proposed to obtain fusion information and then generate original, active, and passive images synchronously. This structure draws inspiration from felines that utilize the unabsorbed light signals through the tapetum lucidum, thereby boosting visual perception capabilities[43-45]. Following the identical principle, the system selectively acquires specific characteristic signals by spatially distributed configuration. The proposed meta-imager integrated with compact bifocal metasurfaces, a low-noise transmitter, a high-sensitivity radiometer, and correlation averaging operations significantly addresses challenges including FOV mismatch, system redundancy, and the lens' aberration. The experiment findings substantiate the viability and precision of fusion information separation imaging, including multi-target active-passive-composite detection, occluded-perception detection, and irregular contour estimation. Moreover, we also installed the APCI on the portable platform for non-interactive hazardous metal detection without any contact, further enhancing the APCI application scope and detecting concealment.

**The Framework of Active-Passive-Composite Aperture-Shared Sub-Terahertz Meta-Imager**

**Fig. 1** illustrates the schematic framework of the proposed APCI, inspired by the adaptive mechanism of the feline visual system. In low-illumination environments, feline tapetum lucidum, the mirror-like microstructure situated behind the retina, selectively redirects incident light back toward the retina, thereby enhancing the response to the specific light, as shown in **Fig. 1a**[46-48]. Drawing inspiration from this biological adaptation, the proposed metasurface-based APCI adopts an active-passive composite imaging architecture, achieving the synchronous acquisition of dual-mode signals, shown in **Fig. 1b**. Among them, metasurfaces possess directional signal manipulation and polarization-dependent response functions. By incorporating a dual-focus design, they achieve a spatially interleaved arrangement of the system's TR components, while cross-polarization isolation of signals on both sides suppresses interference from reflected signals.

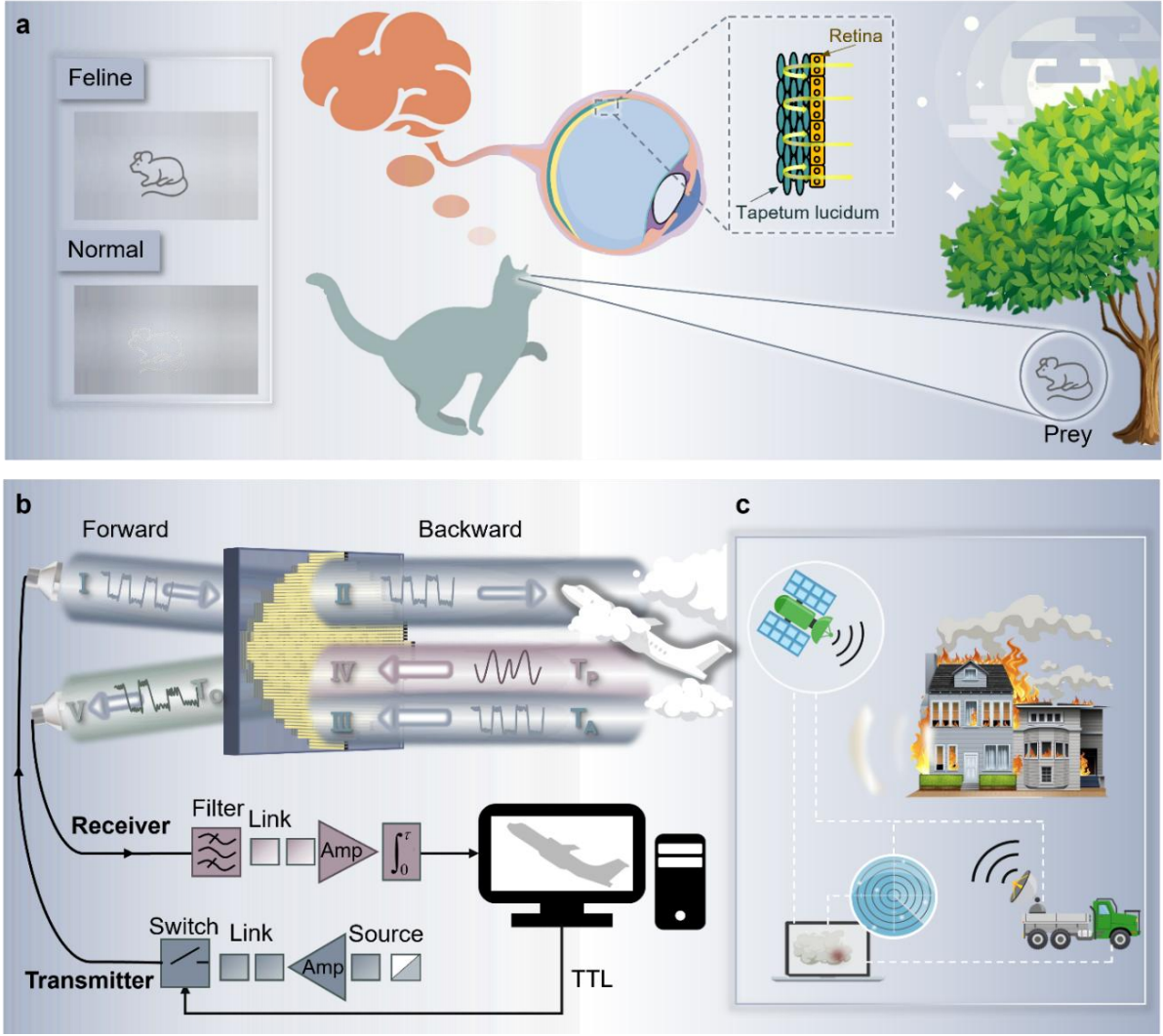

**Fig. 1 Schematic principle of aperture-shared sub-terahertz APCI inspired by feline eyes.**
**a** The tapetum lucidum in feline ocular functions as a retroreflective layer that redirects incident photons back through the retinal layers, doubling the light capture efficiency and thereby significantly enhancing visual sensitivity in dark. **b** Functional demonstration diagram of APCI architecture inspired by the tapetum lucidum. **c** The application prospect of the proposed active-passive-composite aperture-shared meta-imager.

The aperture-shared sub-terahertz APCI utilizes a single radiometer to receive both active and passive signals of the identical frequency band, integrating a dual-mode imaging system at the hardware level. The precise regulation of the metasurface not only significantly enhances the detection range but also greatly improves system integration. Furthermore, the system employs a correlation averaging algorithm to decouple multidimensional fused data, reducing noise and interference while preserving fundamental signal characteristics. By analyzing and averaging correlated data points, the algorithm enhances the accuracy of extracted radiation scattering information. Subsequently, the extracted information is utilized for target feature reconstruction, enabling characterization of the target's shape, physical properties, and material characteristics. In the transmitter section, the system employs a matched load combined with a low-noise amplifier as the core signal source. Following amplification through a power amplifier (PA), the signal is dynamically power-adjusted via an adjustable attenuator, which is governed by a switch and modulated via computer-generated transistor-transistor logic (TTL) signals. The radio frequency signal is ultimately transmitted through an antenna for spatial radiation, thereby completing the information-loaded data emission.

In the active mode, the transmitter emits only preset-modulation signals I, which are subsequently shaped by the metasurfaces into a spatially focused beam II directed toward the observation region. After being contacted with the target, signals III characterizing the target reflection properties are generated, whose energy is denoted as $P_A$,

$$P_A = \frac{P_t \sigma G_t G_r \lambda^2}{(4\pi)^3 r^4} \tag{1}$$

where $P_t$ represents the transmitted power, $\sigma$ stands for the radar cross-section (RCS) of the target, $G_t$ and $G_r$ denote the gains of the transmitting and receiving antennas, $\lambda$ refers to the wavelength of emitted signals, and $r$ indicates the distance between the transmitter and the target.

Concurrently, in passive mode, the target autonomously emits intrinsic EM signatures originating from its thermal-radiative properties (signals IV), denoted as $P_p$,

$$P_P = [eT_0 + (1-e)T_B] \cdot kB \tag{2}$$

where $e$ is the surface's equivalent emissivity, $T_0$ denotes the physical temperature of the target object, $T_B$ refers to the ambient brightness temperature incident on the object, $k$ is the Boltzmann constant, and $B$ is the receiver bandwidth.

Metasurfaces with dual-point focusing integrate signals III and IV and further guide them to form a beam V pointing toward the receiver end, which not only carries the scattering characteristics of targets but also superimposes their radiation features. The radiometer simultaneously acquires active-passive-composite signals via the shared aperture, and the received millimeter wave EM energy is expressed as $P_O$,

$$P_O = P_A + P_P \tag{3}$$

The received signal sequentially passes through a band-pass filter, low-noise amplifier (LNA), a square-law detector, and an integrator to obtain a smooth baseband voltage signal. This processed signal is then segmented based on sampling rate and pixel dimensions, subjected to correlated averaging algorithms, and ultimately output as three-channel images. The calculation process for extracting various feature components from the fused information is described in Section S4 (see Supplementary S4). The active mode performs parameter inversion and feature reconstruction by analyzing reflected echoes containing scattering information. In contrast, the passive detection scheme monitors spontaneous thermal radiation differentials between targets and background environments. As illustrated in **Fig. 1c**, the proposed framework can be deployed across vehicle-mounted and satellite-based multi-platforms to achieve real-time active-passive-composite imaging. It can provide precise perception for both radiation and scattering data, offering robust technical support for anomaly monitoring and emergency rescue operations.

The phase distribution design of metasurfaces applied in the APCI is obtained by the multi-phase superposition method. Here, we set the transmitter located at (0, 50 mm, -300 mm), and emits x-polarized modulated EM waves, forming a y-polarized focus on the detected object side, as depicted in **Fig. 2a**. The backward EM waves, carrying the target features, originate from a point source at (0, 0, 1000 mm) as shown in **Fig. 2b**, and are received at the position (0, -50 mm, -300 mm), which is symmetrical to the transmitter. For the metasurface with the focusing spot located at $(x_i, y_i, f)$, its phase profile is expressed as,

$$\varphi(x, y) = \frac{2\pi}{\lambda} \sqrt{(x-x_i)^2 - (y-y_i)^2 + f^2} - f^2 \tag{4}$$

Overlay three phase profiles of the transmitter $\varphi_{trans}$, receiver $\varphi_{recei}$, and focal point $\varphi_{point}$, to generate the global phase distributions of used metasurface in APCI is obtained,

$$\varphi_{total} = \varphi_{trans} + \varphi_{recei} - \varphi_{point} \tag{5}$$

The optimization of meta-atom arrangement is based on the phase profile distribution of metasurfaces. Detailed structural parameters and simulation outcomes of the meta-atom are provided in Section S1 (see Supplementary S1). **Fig. 2c** illustrates the forward-focusing emulation, achieving an energy conversion efficiency of 80.52% with a full width at half maximum (FWHM) of 18.69 mm, among Airy disk energy accounting for 81.22%. The peak intensity occurs at (2 mm, 0, 978 mm). A comprehensive analysis of spatial propagation losses and other irrelevant components is provided in Section S2 (see Supplementary S2). In the reverse process, the total energy conversion efficiency reaches 88.98%. The longitudinal focal length is measured 295 mm, with FWHM values of 5.44 mm and 5.28 mm, respectively. In addition, robust performance under non-ideal conditions is verified (see Supplementary S2), including variations in incident angles and shifts in operating frequency. **Fig. 2g** displays metalens fabricated by PCB technology and inset local zoom area. The experimental setup and measurement process are detailed in **Fig. S7** (see Supplementary S3). Under front and back incidence, **Fig. 2e, f** demonstrates the experimentally measured intensity distributions, which demonstrate strong consistency with corresponding simulations at an operational frequency of 94 GHz. **Fig. 2h-j** present magnified views of individual focal spots and the normalized energy distribution measured along the centerline of the focal points. After disregarding low-value noise bias, the experimental profiles align with numerical simulations in both trend and magnitude, which demonstrate exceptional field confinement capabilities. The slight energy deviation around the focus may result from manufacturing errors in the metalens.

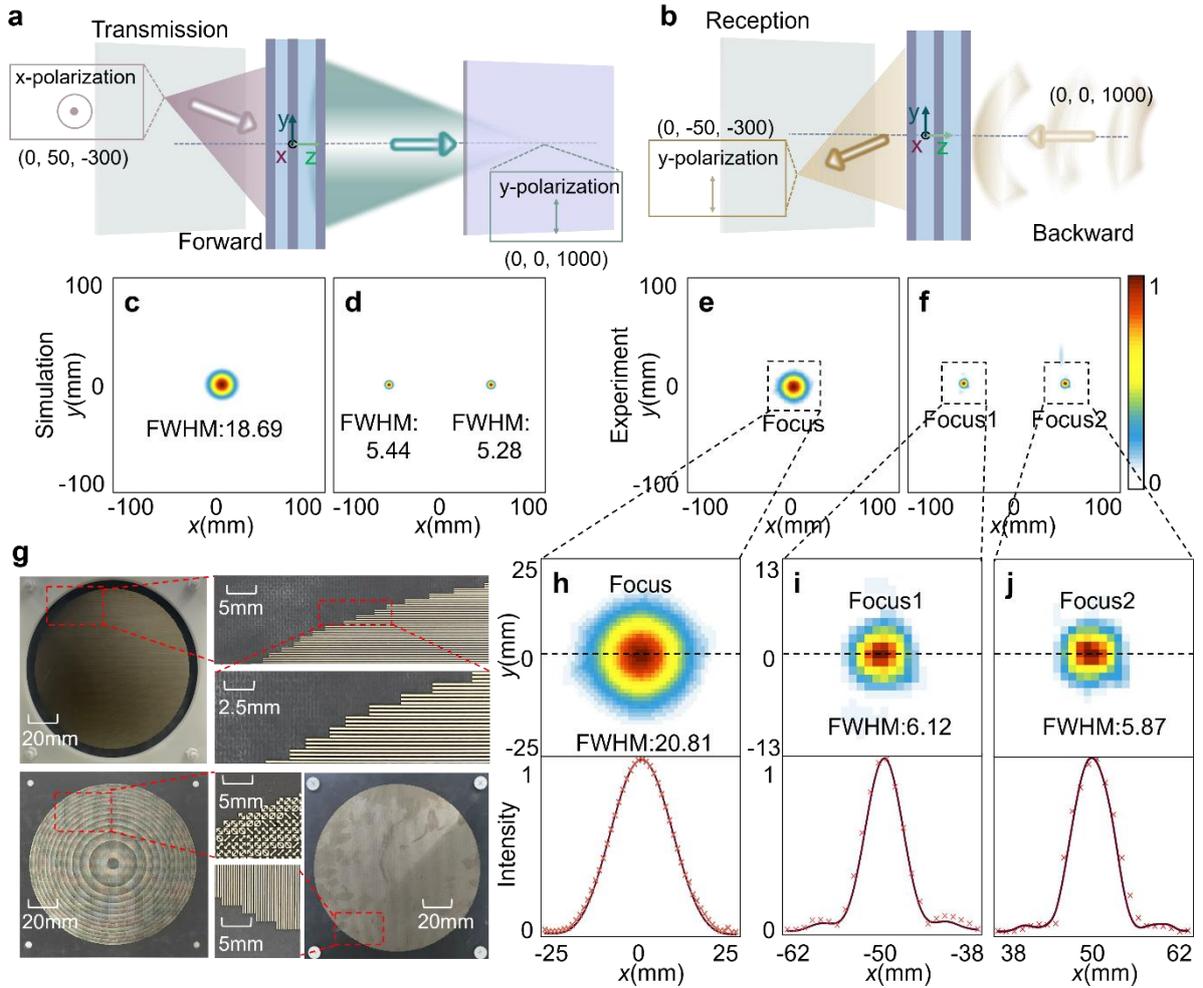

**Fig. 2 Schematic diagram of the bifocus metasurfaces for shared transmitting-receiving aperture. a** Signals emitted from the transmitter establish a cross-polarization focal point. **b** Signals emitted from the backward generate two identical polarization focal points, one of which serves as the acquisition point. **c-d** Simulation diagram of the energy distribution under forward and backward emission. **e-f** Measured energy distribution for forward and backward processes. **g** Optical image of the fabricated prototype. **h-j** Local amplification and normalized intensity curve of measured Focus, Focus1, and Focus2.

## Results and Discussion

**Fig. 3a** illustrates the simulated target detection scenario of APCI under modulated signal illumination, along with the system's signal flow diagram. The signal source generates random signals, which are modulated and then output as $V_{out}$. The received signal $V_{in}$ contains environmental background noise and target feature parameters. It undergoes demodulation processing to produce the final output waveform. When the system scans a strongly scattering object, the received signal level increases significantly, whereas, for a strongly radiating object, the signal fluctuation amplitude exhibits a linear relationship with its brightness temperature.

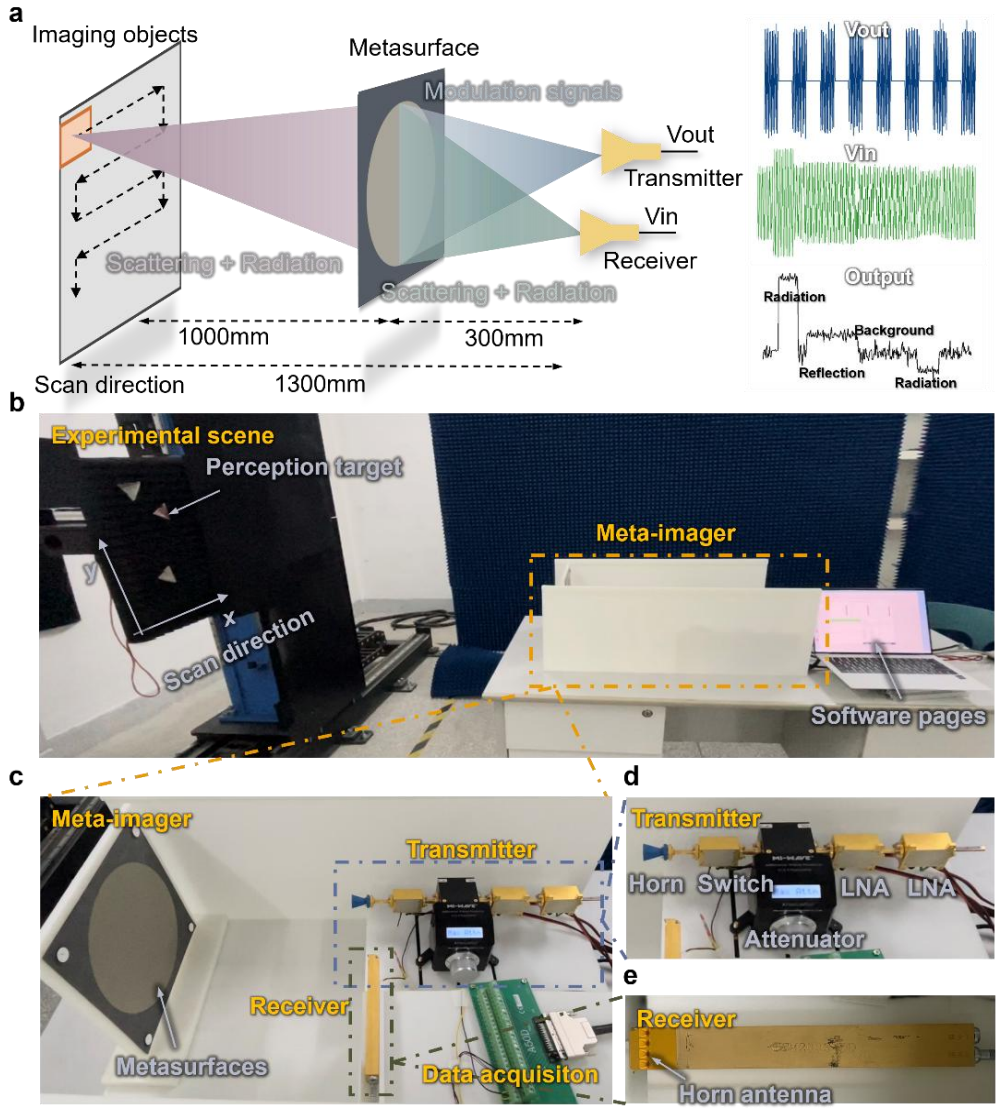

**Fig. 3 Experimental setup of the active-passive-composite aperture-shared meta-imager in the sub-terahertz frequency band. a** Imaging process and signal models. **b** Experimental scenario. Data separation and image display are handled on the computer. **c** Imaging system. **d** Transmitter. Signal modulation is implemented by the computer. **e** Receiver.

The system architecture, depicted in **Fig. 3b**, integrates a dual-axis scanning stage, metasurface modules, and encapsulated imaging components, with real-time visualization of reconstructed images. APCI employs resin encapsulation technology and consists of four core subsystems, transmission assembly, metalens, reception module, and data processing system, as detailed in **Fig. 3c**. The transmitter, configured with 60 dB attenuation, yielding an equivalent radiated power of -60 dBm, integrates key parts as shown in **Fig. 3d**. As shown in **Fig. 3e**, the radiometer as the receiver captures radiation-scattered signals, supporting both active and passive detection modes. Although the radiometer features four operational channels, only one was utilized during the experiments. Comprehensive descriptions of system detection principles and critical component specifications are available (see Supplementary S4). During imaging, the stationary imaging module achieves spatial scanning through coordinated movement of the displacement platform. Received signals were used for subsequent active-passive fusion processing, ultimately reconstructing brightness temperature distributions. During image processing, Gaussian filters were employed to denoise acquired data, thereby improving the spatial resolution and feature fidelity. The system provides an imaging resolution of approximately 24 mm, with the radiometer sensitivity exceeding 0.5 K.

Preliminary experiments were conducted in a controllable indoor environment with a scanning range of 1.5 m × 1.5 m. Distinct target objects were designed for active and passive modes to verify the dual-modal detection capabilities. The active detection mode focused on metallic objects with strong EM wave reflectivity, whereas the passive mode identified non-metallic subjects exhibiting significant thermal variations or material contrasts. Three types of standardized test pieces were used, metal angle reflectors with a side length of 30 mm, a square heating plate with dimensions of 100 mm × 100 mm, and a rectangular cooling plate measuring 100 mm × 150 mm. They were arranged as shown in **Fig. 4a**, with heat and cold sources obscured by black microwave-absorbent material to attenuate their radiation signatures. The ambient temperature was maintained at approximately 300 K, while the heating plate raised local temperatures above 345 K, and the cooling plate reduced them below 290 K, creating pronounced thermal gradients relative to the background.

The original results, as presented in **Fig. 4b**, exhibited more pronounced brightness temperature anomalies in the area of covered heat and cold pads compared to the reflectors under weak excitation conditions. The experimental images showed that APCI had achieved the separation of multi-target composite information, including scattering information of metal targets and radiation information of heating-cooling plates, as shown in **Fig. 4c, d**. Despite identical geometric parameters, variations in reflector orientation induced significant discrepancies in reflected intensity, leading to inconsistent brightness temperature. As shown in **Fig. 4e**, the region containing the metallic corner reflector was subjected to heating. Due to the near-zero emissivity of metallic materials, they cannot be undetectable in passive imaging even when heated. **Fig. 4f–h** further demonstrate that, against a background with significant radiative brightness temperature variations, the scattering characteristics of metallic targets could still be distinctly extracted, enabling effective identification of hidden targets.

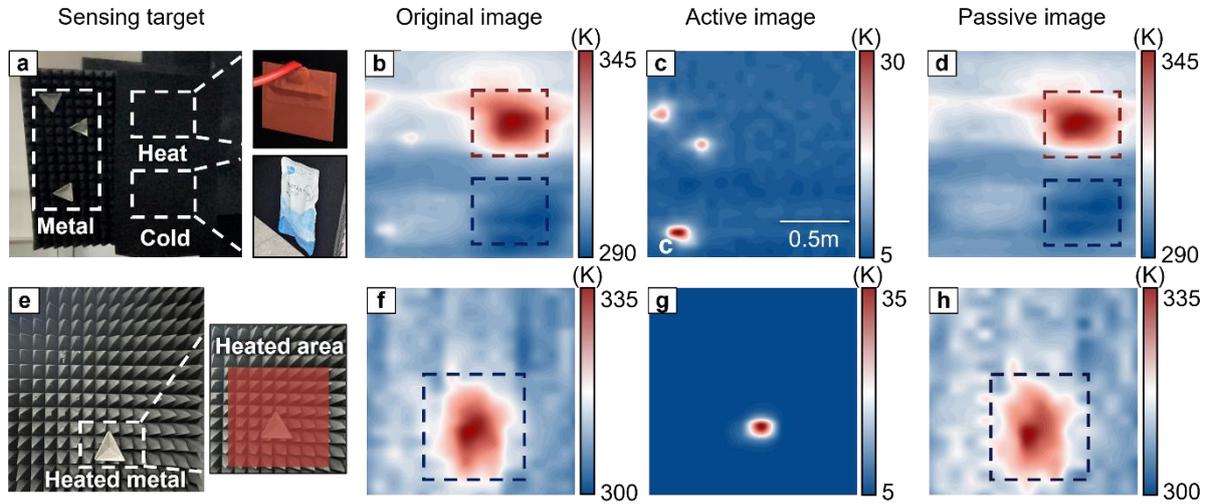

**Fig. 4 Multi-source fusion information imaging verification. a-d** Experimental target setup and fusion-image separation. On the left, three identical metal reflectors were randomly placed, while on the right, heating and cooling plates were covered. **e-h** Experimental setup and separated images. The heating centered around the metal.

APCI integrates both passive and active imaging to capture the radiative and scattering properties of targets, enabling comprehensive extraction of high-dimensional feature information. This active-passive-composite detection approach not only enhances detection accuracy but also provides feasible solutions for non-interactive security screening. This system is capable of autonomous, covert checks without active subject cooperation, which is particularly advantageous in addressing extreme threats, such as suicide bombings. Traditional checks rely on direct physical contact, which not only increases the risk of human intervention but also provokes terrorists, exacerbating security threats. In contrast, APCI employs a portable dual-axis scanning platform for non-contact detection and utilizes camouflage techniques to enable non-cooperative screening, as illustrated in **Fig. 5a**. Suicide bombs typically consist of explosives and a large amount of metal screws and fragments, designed to maximize the blast radius and destructive power upon detonation. To simulate such devices, the system targets these metals for detection while also conducting identification experiments on dangerous items such as knives and firearms, as shown in **Fig. 5b**. The introduced radiation-scattering-joint feature distribution provides a quantitative analysis of detection results. As shown in **Fig. 5c**, the scenarios of only humans and those involving various carried objects form distinct and independent clustering regions. The scattering and radiation characteristics of different targets exhibit significant variations, indicating that the reflectivity of an object plays a key role in its scattering properties and affects human radiation data. Furthermore, whether detecting a human body alone or one carrying an explosive vest, the distribution remains concentrated in the same region when observed from different directions, as shown in **Fig. 5d**. This suggests that directional changes have minimal influence. These findings validate the preliminary classification capability of non-imaging detection, enabling autonomous, concealed, and non-interactive security screening in public environments. Notably, the system is expected to leverage an adaptive spatial sampling mechanism to support high-resolution near-field inspection and rapid far-field screening, further enhancing the efficiency and adaptability of security checks.

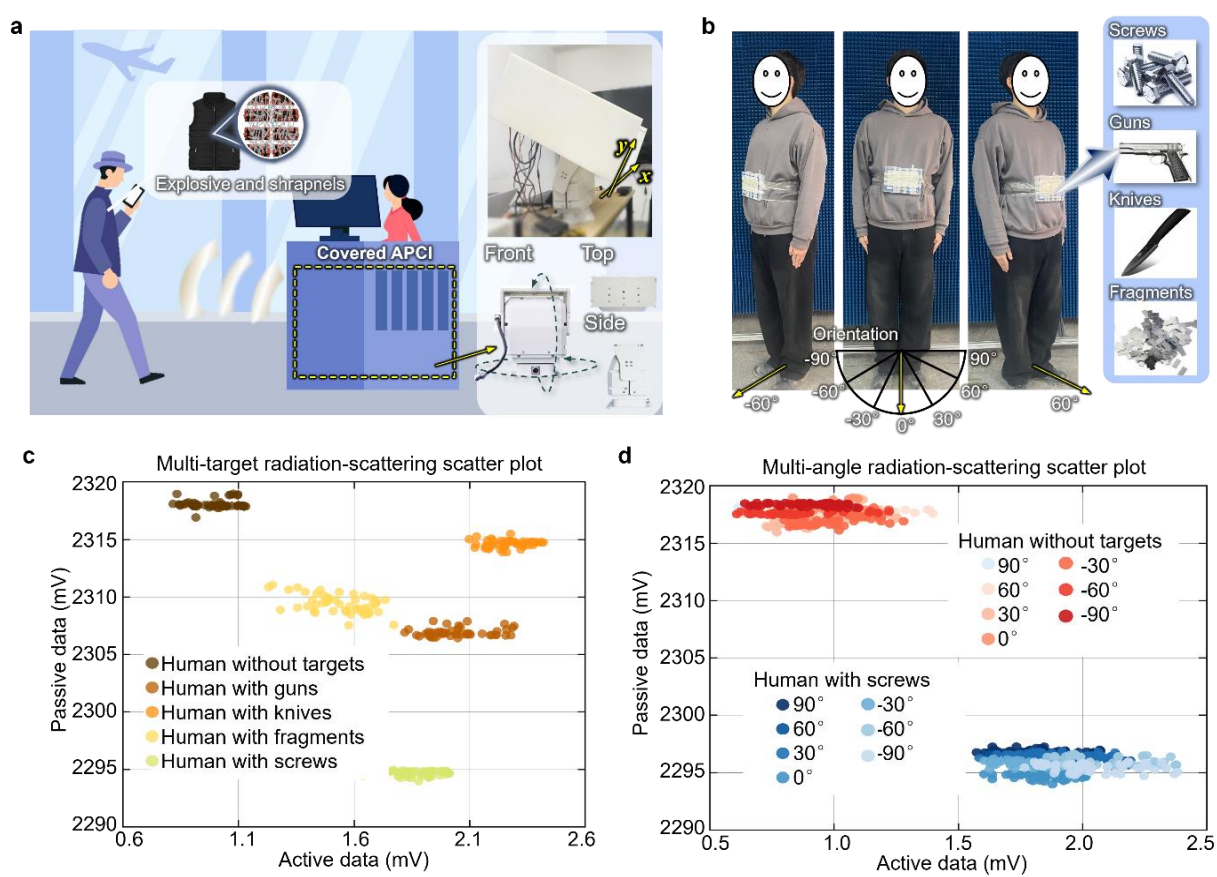

**Fig. 5 Non-interactive concealed metals detection with the proposed portable APC imager. a** Concealed non-contact detection scenarios. Disguised APC imagers integrate a dual-degree-of-freedom portable platform, enabling non-interactive detection of explosive devices carried by terrorists. **b** Visible optical image of the detection scene with invisible metals. **c** Scatter plot of radiation and scattering data for different targets. **d** Scatter plot of radiation and scattering data for multi-orientation targets.

Sub-terahertz APC imagers fulfill the demand for miniaturization and integration, owing to the metasurfaces' flexible and powerful regulation capability (see Supplementary S5). Another standout feature of the meta-imager is its simplified hardware structure. In sharp contrast to an additional reflector or large-scale array reception antenna, it achieves single-pixel detection through only a single-channel transmitter and receiver without extra components. **Table 1** compares critical performance metrics including compactness, flexibility, and detection sensitivity, demonstrating that APCI achieves functional integration and an expanded detection range. Furthermore, APCI outperforms both single-modality systems and non-metasurface systems in multi-target detection tasks, which exhibits robust imaging performance across the entire W-band (see Supplementary S6).

It should be noted that the metasurface design emphasizes near-field focus to facilitate indoor operations. However, its flexibility is not restricted to near-field applications. They can also be customized to adjust the FOV and detection distance, which provides unprecedented convenience for the system's application, enabling the system to function effectively in complex environments and meet diverse operational requirements. Additionally, irregular-shaped target detection was performed, and the outline and orientation of the target were roughly determined in the active image (see Supplementary **Fig. S13**).

**Table 1.** Comparison of the APCI with other same-type systems.

| Ref. | Type | Multi-target or not | Freq. (GHz) | Detection Range(m) | Res. (m) | Temp. Res.(K) | System Size($m^3$) |
|---|---|---|---|---|---|---|---|
| 15 | Active-passive composite | No | 116-131(Act.) 94(Pass.) | 1.7 | 0.03 | 0.05 | About 1×1×1.5 |
| 18 | Active-passive composite | No | 170-220(Act.) 35(Pass.) | 0.2-0.3(Act.) 1.45-1.7(Pass.) | 0.03 | N/A | N/A |
| 19 | Active-passive composite | No | 32-36 | 4 | 0.04 | 2 | 1.6×1×2 |
| 52 | Active | Yes | 38 | 1.83 | 0.02 | N/A | About 0.6×1.2×0.2 |
| 53 | Passive | Yes | 27-40 | 4 | 0.04 | 2 | 1.6×0.8×0.5 |
| **This work** | **Active-passive composite** | **Yes** | **75-110** | **0.8** | **0.024** | **0.5** | **0.71×0.415 ×0.285** |

Ref., reference; Freq., frequency; Res., resolution; Act., active; Pass., passive.

**Discussion**

In conclusion, we propose the aperture-shared sub-terahertz APCI based on metasurfaces that mimics the feline tapetum lucidum, achieving the composite of active and passive detection systems. Through spatial interleaving layouts and correlation averaging processing, active scattering and passive radiation data extractions in the sub-terahertz band are implemented, offering a solution for single-region information decoupling and multi-target fusion imaging.

This highly integrated and high-throughput sub-terahertz APCI drives the evolution of detection systems toward miniaturized deployment and real-time data readout. In addition, we also introduce the non-imaging security screening scheme. Taking the human-alone radiation-scattering value as the baseline reference, the approach enables non-interactive autonomous identification and covert monitoring of concealed objects in public spaces. More importantly, the proposed device is easily scalable across the spectrum, thus providing significant reference value for imaging applications from microwaves to optics.

## Methods

**Numerical simulation**

In the forward design process, the finite-difference time-domain (FDTD) method in the commercial software CST was used to numerically simulate the amplitude and phase response of unit cells with varying widths and rotation angles. In the final metasurface arrangement, only 45° and 135° rotations were employed. Detailed structural parameters and simulation results of the unit cells are provided in Supplementary Note 1.

**Fabrication**

The proposed metasurfaces in this study were fabricated using a printed circuit board (PCB) manufacturing process. They consist of three metallic layers and two dielectric layers. The dielectric material is RT5880, with a thickness of 0.508 mm and a relative permittivity of 2.2. The metallic layers have a thickness of 0.018 mm, and the unit cell periodicity is 1 mm. The final metasurface retains only the unit cells within a 100 mm radius.

**Data availability**

The relevant data and images in this article can be viewed in the main text and supplementary information, and the original data can be obtained from the relevant authors as required.

**Code availability**

The training code for this article can be obtained from relevant authors upon request.

**Acknowledgements**

This work was supported in part by the National Natural Science Foundation of China under Grant 62271170 and Grant 62371159, National Key Laboratory of Laser Spatial Information Foundation, the Natural Science Foundation of Heilongjiang Province under Grant YQ2023F007, the Fundamental Research Funds for the Central Universities under Grants FRFCU5710052821, and the assisted project by Heilong Jiang Postdoctoral Funds for scientific research initiation under Grant LBH-Q21093.

**Author contributions**

M. H., Y. Z. W., and Z. J. contributed equally to the work. M. H. and Y. Z. W. conceived this idea and achieved the experimental design. Z. J., Y. L., and Z. S. conducted simulations and validations. Z. Y., Y. D. L., and P. W. designed and assembled the measurement setup. Y. F. W., A. Y., and Z. K. assisted in conducting experiments, while P. C., Y. H., and W. L. carried out the theoretical analysis. M. H., Y. Z. W., Y. C., and J. Q. wrote the manuscript. Y. Y., J. Q., and Y. D. provided effective result analysis and evaluation and supervised the project. All authors have made significant contributions to the writing of the paper.

**Competing interests**

The authors declare no competing interests.